\documentclass {article}
\usepackage{spconf,amsmath,graphicx}
\usepackage{url,subfigure,cite,array}
\usepackage{textcomp}
\usepackage{stfloats}
\usepackage{hyperref} 
\usepackage{subfigure}
\usepackage{verbatim}
\usepackage{makecell}
\usepackage{multirow}
\usepackage{graphicx}
\usepackage{cite}
\usepackage{amssymb}
\usepackage{enumerate}
\usepackage{color}
\usepackage{xcolor}
\usepackage{amsmath,amsfonts}
\usepackage{algorithmic}
\usepackage{algorithm}
\usepackage{array}
\usepackage{subfigure}
\usepackage{makecell}
\title{Delivering Speaking Style in Low-resource Voice Conversion with Multi-factor Constraints}
\name{Zhichao Wang$^1$, Xinsheng~Wang$^1$, Lei Xie$^1$$^*$, Yuanzhe Chen$^2$, Qiao Tian$^2$, Yuping Wang$^2$}

\address{
  $^1$Audio, Speech and Language Processing Group (ASLP@NPU)\\School of Computer Science,
  Northwestern Polytechnical University, Xi’an, China\\
  $^2$Speech, Audio \& Music Intelligence (SAMI), ByteDance}

\begin{document}
\ninept
\maketitle
%

\begin{abstract}
Conveying the linguistic content and maintaining the source speech's speaking style, such as intonation and emotion, is essential in voice conversion (VC). However, in a low-resource situation, where only limited utterances from the target speaker are accessible, existing VC methods are hard to meet this requirement and capture the target speaker's timber. In this work, a novel VC model, referred to as MFC-StyleVC, is proposed for the low-resource VC task. Specifically, speaker timbre constraint generated by clustering method is newly proposed to guide target speaker timbre learning in different stages. Meanwhile, to prevent over-fitting to the target speaker's limited data, perceptual regularization constraints explicitly maintain model performance on specific aspects, including speaking style, linguistic content, and speech quality. Besides, a simulation mode is introduced to simulate the inference process to alleviate the mismatch between training and inference. Extensive experiments performed on highly expressive speech demonstrate the superiority of the proposed method in low-resource VC.

\end{abstract}

\renewcommand{\thefootnote}{\fnsymbol{footnote}}
\footnotetext{*Corresponding author.}

\section{Introduction}

Voice conversion (VC) aims to modify speech from a source speaker to sound like that of a target speaker without changing the linguistic content and speaking style~\cite{imliu2020transferring}. Early related work mainly focused on the speaker identity conversion but paid limited attention to the style of source speech~\cite{VAEHsu2016Voicevae,PPGSun2016PhoneticPF,GANHsu2017VoiceCF}. However, maintaining the source speech's speaking styles, i.e., emotion and intonation, is important in many scenarios, such as dubbing, live broadcasting, and data augmentation. 
\begin{figure*}[ht]
\centering
\begin{minipage}{0.55\linewidth}
    \subfigure[MFC-StyleVC]{
      \includegraphics[width=1\columnwidth]{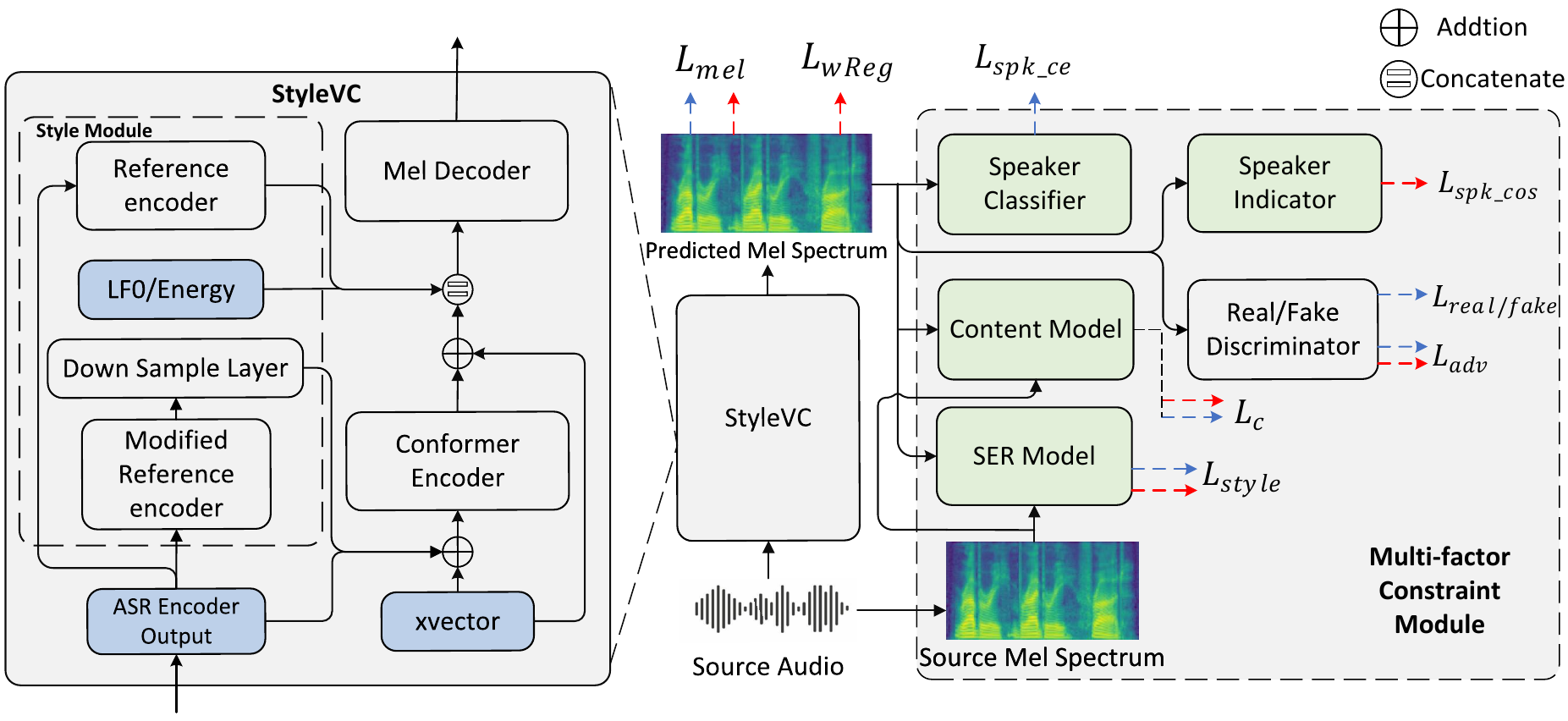}}
\end{minipage}
\begin{minipage}{0.4\linewidth}
    \subfigure[Speaker Indicator]{
      \includegraphics[width=.95\columnwidth]{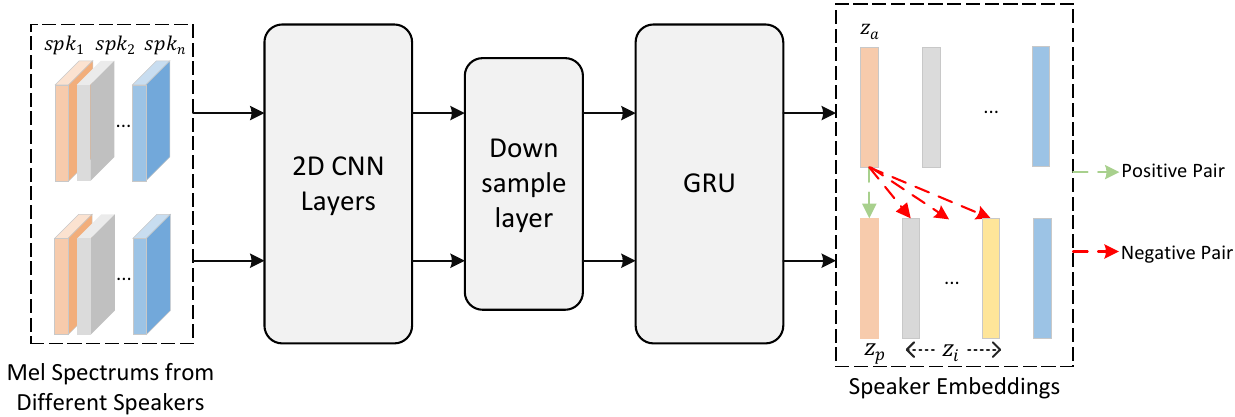}}\\
    \subfigure[SER]{
      \includegraphics[width=.95\columnwidth]{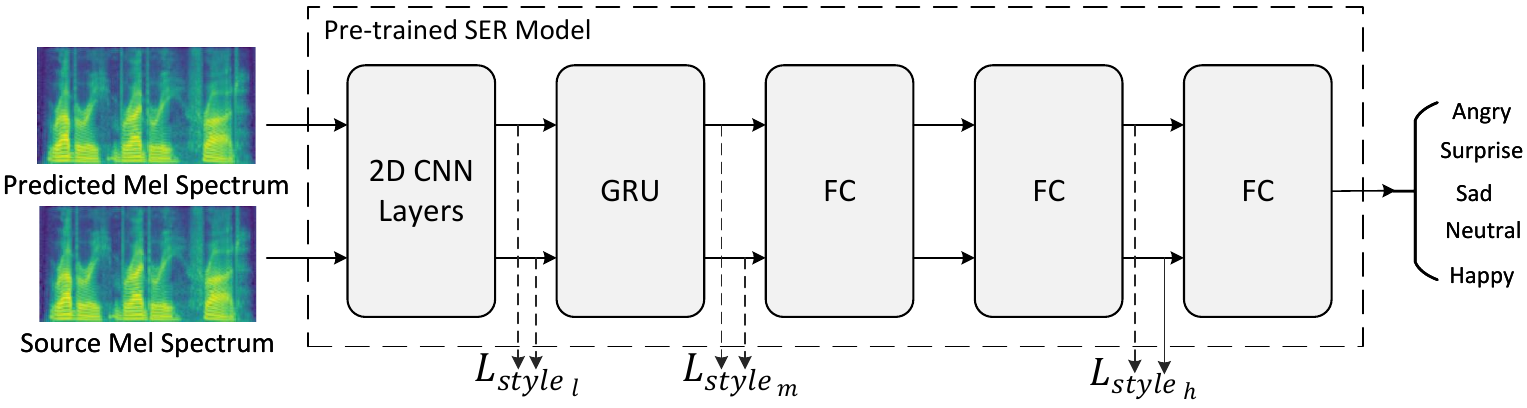}}
\end{minipage}

\caption{The overall architecture for MFC-StyleVC. (a). The MFC-StyleVC model. (b). The speaker indicator model. (c). The SER model. Note that in (a), the blue and red dashed arrows represent the objective function for base model training and adaptation, respectively. The block in the green means a pre-trained model which is fixed in the whole process.}
\label{fig:mfc-stylevc}
\end{figure*}
While some recent effort has been conducted toward better source speaking style maintenance with prosodic features ~\cite{explicitMing2016ExemplarbasedSR,explicitRaitio2020prosodyControl} or style embeddings extracted from the style extractor~\cite{imdu2021expressivevc,imgan2022iqdubbing,imLian2021TowardsFP,imliu2020transferring,imwang2021enriching}, these methods generally rely on a large number of recordings, which is hard to meet the demand with limited speech recordings. 

The main challenge of the VC task with limited recordings from the target speaker is the over-fitting issue, which makes it hard to balance high-quality speech generation, speaker identity conversion, and style delivery. To obtain a VC model for the case with extremely limited target speech, i.e., only one utterance from the target speaker is accessible, which is also called \textit{one-shot VC}, speaker adaption is adopted in~\cite{wang2022one}, in which process the parameters of a pre-trained VC model is further updated based on the accessible utterances of the target speaker. While some strategies, e.g., speaker normalization and weight regularization, are proposed in~\cite{wang2022one} to help the adaptation process, the issue of style delivery in highly expressive scenarios is not explicitly taken into consideration. In contrast, obtaining the speaker representation from an utterance of the target speaker with a well-trained speaker embedding extractor is another popular strategy~\cite{autovcqian2019autovc,INchou2019oneshot,OSSEDu2021ImprovingRO,VQMIVC,yin2022retriever}, in which the parameters of the well-trained VC model are not further updated on the target speaker, preventing the model from over-fitting to the given utterance of the target speaker. However, with limited information conveyed by the speaker embedding and the lack of targeted optimization for the target speaker, the ability of such methods on speaker identity conversion is limited, leading to low speaker similarity.

To face these difficulties, MFC-StyleVC, which is a speaker adaptation-based VC framework for delivering speaking style to a low-resource speaker, is proposed in this work. Specifically, obtaining high speaker timbre similarity to the target speaker is an essential goal of the low-resource VC task. When no explicit guidance exists toward the target speaker timbre in the adaptation process, it is hard to achieve desired performance due to the limited utterances from the target speaker. The speaker timbre constraint, which is provided by a speaker classifier and speaker indicator, is utilized to help the model in capturing the target speaker timbre in base model training and adaptation. With limited generalization on low-resource unseen speakers for speaker classifier, a contrastive learning-based speaker indicator is newly proposed to guide the speaker adaptation process. However, if only the speaker timbre representation of the generated result is forced to be similar to that of target speaker without other constraints, the adapted model would be over-fitted to the limited content. Therefore, perceptual regularizations, including speaking style, content, and speech quality constraints, are adopted to maintain the abilities learned in base model training and make the model focus on adapting to new speakers.
Inspired by the training process of meta-learning~\cite{min2021meta}, a simulation mode in training is introduced to reduce the mismatch between training and inference.
Experiments demonstrate that MFC-StyleVC achieves superior performance to the previous state-of-the-art systems on speaker adaptation and speaking style delivery. The effectiveness of each constraint is proved by the ablation study, which indicates the good design of the proposed model.

\section{Proposed methods}

As shown in Fig.~\ref{fig:mfc-stylevc}a, the proposed framework mainly consists of StyleVC and a multi-factor constraint module. The StyleVC is comprised of a style module, a conformer encoder, and a mel decoder. The style module in StyleVC is to obtain global, local~\cite{muti-scale}, and frame-level style representations from ASR encoder output and prosodic features of source speech, including logarithmic domain fundamental frequency (lf0) and short-term average amplitude (energy). The conformer encoder uses speaker representation xvector, ASR encoder output, and local-level style representation as input. The mel decoder predicts mel spectrum from xvector, conformer output, global style, and frame-level prosodic features. During training process, the multi-factor constraint module generates model's constraints, including the speaker timbre, speaking style, linguistic content, and speech quality, from mel spectrums of converted and natural speeches. The whole training process consists of training base model and adapting to target speaker and different constraints involve different processes. With the generated spectrum, a universal TFGAN~\cite{tian2020tfgan} is adopted to reconstruct waveform. The proposed multi-factor constraints and training strategy will be detailedly introduced in this section.

\vspace{-3pt}
\subsection{Speaker Timbre Learning}
Obtaining high speaker similarity to the target speaker is an essential goal of the VC task. However, it is non-trivial to achieve this goal in the low-resource scenario, because limited utterances from the target speaker can be used. It is effective to add an external constraint to guide the model to adapt to target domain with limited utterances and find the optimal parameters\cite{surveywang2020generalizing}. Therefore, the speaker timbre constraint is proposed to explicitly constrain the speaker timbre information via a speaker classifier and a speaker indicator. 

\textit{1) Speaker classifier:} During the base model training process, sufficient recordings from each speaker make the speaker classifier loss a possible way to constrain the speaker timbre. Therefore, a speaker classifier is introduced to guide the learning of speaker timbre. In practice, the speaker classifier takes the predicted mel spectrum as input and outputs the probability of this spectrum belonging to the current speaker identity. The classification loss is obtained by the cross entropy, which is referred to as $\mathcal{L}_{spk\_ce}$.

\textit{2) Speaker indicator:} Although the speaker classification loss is intuitive for the speaker constraint, it is unsuitable for the low-resource target speaker during adaptation. Poor generalization and insufficient recordings limit the use of classifier. Therefore, similar to Wan et al.~\cite{SV}, another speaker indicator is proposed to constrain the similarity between speaker embeddings. The basic idea of this speaker indicator is to increase the similarity of speaker embeddings from the same speaker and decrease the similarity of speaker embeddings from different speakers. As shown in Fig.~\ref{fig:mfc-stylevc}b, the architecture of the speaker indicator is similar to the reference encoder that takes the predicted spectrum as input and outputs the speaker embedding. A triplet loss is adopted to make sure the low inter-speaker similarity and high intra-speaker similarity. To be specific, with a speaker embedding $z_a$ produced by the speaker indicator as an anchor, another speaker embedding with the same speaker identity as the positive sample $z_p$, and k embeddings, referred to as $z_i$, with different speaker identities that randomly sampled within the batch as negative samples, the objective function can be defined as:
\begin{equation}
    \begin{split}
    \mathcal{L}_{spk\_indcator}=(1-\frac{z_{a}\cdot z_{p}}{||z_{a}||^{2}\cdot||z_{p}||^{2}}) + \frac{1}{k}\sum_{i}^{k}\frac{z_{a}\cdot z_{i}}{||z_{a}||^{2}\cdot||z_{i}||^{2}}.
    \end{split}
\end{equation}
With the help of the speaker indicator, the speaker similarity loss between speaker embedding $\hat{z}_{s}$ of predicted mel spectrum and target speaker embedding $z_s$ can be defined as:
\begin{equation}
\mathcal{L}_{spk\_cos}=1-\frac{\hat{z}_{s}\cdot z_{s}}{||\hat{z}_{s}||^{2}\cdot||z_{s}||^{2}}
\end{equation}

\begin{table*}[]
\scriptsize
\centering
\caption{Comparison of different models. Note that $2.97$ and $0.886$ are calculated from testdata and target speaker data.}
\vspace{5pt}
\label{exp:mos&obj}
\begin{tabular}{c|cc|cc|cc|cc|cc|cc|cc}
\hline
             & \multicolumn{2}{c|}{Speech Quality $(\uparrow)$}  & \multicolumn{2}{c|}{\makecell[c]{Speaker Similarity $(\uparrow)$}}
             & \multicolumn{2}{c|}{StyleCMOS $(\downarrow)$}
             & \multicolumn{2}{c|}{\makecell[c]{CER\\$(\downarrow, 2.97)$}}         & \multicolumn{2}{c|}{$D_{style} (\downarrow)$}     &\multicolumn{2}{c|}{$P_{LF0} (\uparrow)$}               &\multicolumn{2}{c}{\makecell[c]{Cos.Sim\\$(\uparrow, 0.886)$}}        \\ \hline
Length      & \multicolumn{1}{c|}{1 utt.} & 5 utt. & \multicolumn{1}{c|}{1 utt.} & 5utt & \multicolumn{1}{c|}{1 utt.} & 5utt & \multicolumn{1}{c|}{1 utt.} & 5 utt. & \multicolumn{1}{c|}{1 utt.} & 5 utt. & \multicolumn{1}{c|}{1 utt.}  & 5 utt.    & \multicolumn{1}{c|}{1 utt.}  & \multicolumn{1}{c}{5 utt.}  \\ \hline
Hybrid-VC   & \multicolumn{1}{l|}{3.34$\pm$0.12} & 3.65$\pm$0.12 & \multicolumn{1}{l|}{3.37$\pm$0.14} & 3.43$\pm$0.12 &\multicolumn{1}{l|}{-0.619} &-0.193 & \multicolumn{1}{l|}{4.63} & 3.85 & \multicolumn{1}{l|}{3.18} & 2.38 & \multicolumn{1}{l|}{0.613} & 0.656  & \multicolumn{1}{l|}{0.798} & \multicolumn{1}{l}{0.815} \\
SN-VC       & \multicolumn{1}{l|}{3.28$\pm$0.09} & 3.75$\pm$0.10 & \multicolumn{1}{l|}{3.35$\pm$0.10} & 3.59$\pm$0.13 &\multicolumn{1}{l|}{-0.952}  &-0.726  & \multicolumn{1}{l|}{7.86} & 5.73 & \multicolumn{1}{l|}{3.21} & 2.64 & \multicolumn{1}{l|}{0.593} & 0.681  & \multicolumn{1}{l|}{0.802} & \multicolumn{1}{l}{0.826} \\
StyleVC     & \multicolumn{1}{l|}{3.42$\pm$0.09} & 3.75$\pm$0.10 & \multicolumn{1}{l|}{3.41$\pm$0.11} & 3.55$\pm$0.11 &\multicolumn{1}{l|}{-0.297}  &-0.285   & \multicolumn{1}{l|}{4.41} & 3.86 & \multicolumn{1}{l|}{3.41} & 2.77 & \multicolumn{1}{l|}{0.627} & 0.644  & \multicolumn{1}{l|}{0.796} & \multicolumn{1}{l}{0.818}\\
MFC-StyleVC & \multicolumn{1}{l|}{\textbf{3.70$\pm$0.10}} & \textbf{3.80$\pm$0.12} & \multicolumn{1}{l|}{\textbf{3.65$\pm$0.11}} & \textbf{3.75$\pm$0.11} &\multicolumn{1}{c|}{-} &- & \multicolumn{1}{l|}{\textbf{3.98}} & \textbf{3.74} & \multicolumn{1}{l|}{\textbf{2.26}} & \textbf{2.16} & \multicolumn{1}{l|}{\textbf{0.681}} & \textbf{0.722}  & \multicolumn{1}{l|}{\textbf{0.804}} & \multicolumn{1}{l}{\textbf{0.828}} \\ \hline
\end{tabular}
\end{table*}

\vspace{-10pt}
\subsection{Perceptual Regularization Constraints}

With only speaker timbre constraint forcing the learning of new speaker timbre during adaptation, 
the adapted model would be over-fitting to the limited utterances. It is necessary to prevent the abilities of the model learned in base model training from being forgotten, including style delivery, content preservation, and high-quality generation. The regularization constraints on these aspects are employed to achieve the purpose and make the model parameters focus on being optimized in the speaker timbre space during adaptation.

\textit{1) Speaking style consistency:} To prevent the model from over-fitting the speaking style of the target speaker in the low-resource case, a speech emotion recognition (SER)~\cite{liurui2021expressive} model is used to calculate style matching loss. This regular term is calculated from different levels, which stand for the abstraction degree of the hidden representation to the style. As shown in Fig.~\ref{fig:mfc-stylevc}c, features from different hidden layers indicated as $h_{l}$, $h_{m}$, and $h_{h}$, respectively, are used to obtain the style matching loss:
\begin{equation}
    \mathcal{L}_{style_{s}}=||h_s-\hat{h_s}||^{2}_{2},s\in \{ l, m, h \}
\end{equation}
where $h$ and $\hat{h}$ are extracted from the SER model using source mel spectrum and predicted mel spectrum as input.


\textit{2) Linguistic content:} A content loss generated by a content model ensures pronunciation stability during adaptation. With the same architecture as CBHG module\cite{TacotronWang2017}, in the pre-training process, this content model uses mel spectrum as the input, and its prediction object is the content feature extracted from ASR encoder. During speaker adaptation, the content loss between content features $c$ and $\hat{c}$ extracted from the source mel spectrum and predicted mel spectrum can be described as:
\begin{equation}
\mathcal{L}_{c}=||c-\hat{c}||^{2}_{2}
\end{equation}

\textit{3) Speech quality:} A real/fake discriminator $D$ optimized in an adversarial manner is adopted for preventing speech quality degradation. As shown in Fig.~\ref{fig:mfc-stylevc}, the real/fake discriminator takes mel spectrum as input to judge whether the spectrum is real or not. The discriminator is trained to be confused on distinguishing real spectrum $y$ and generated spectrum $\hat{y}$. This process can be described as:
\begin{equation}
\mathcal{L}_{real\_fake}=||1-D(y)||^{2}_{2}+||D(\hat{y})||,
\end{equation}
\begin{equation}
\mathcal{L}_{adv}=||1-D(\hat{y})||^{2}_{2}.
\end{equation}
During adaptation, the parameters of the real/fake discriminator are fixed, and only $L_{adv}$ is used to optimize the VC model. 

\vspace{-3pt}
\subsection{Training Strategy: Reconstruction and Simulation}
In addition to the losses mentioned above, mel reconstruction loss $\mathcal{L}_{mel}$ and weight regularization loss $\mathcal{L}_{wReg}$~\cite{weightRegLi2020,wang2022one} are introduced to ensure the reconstruction of mel spectrum and further alleviate over-fitting. Note that $\mathcal{L}_{wReg}$ is a variant of l2 regularization to prevent the parameters from drifting far away from the base model. Besides, in the inference stage, the speaker timbre, speaking content,
and speaking style are not from the same speech, which is different from the training process, and thus leads to potential performance degradation in inference. To alleviate this issue, inspired by meta-learning\cite{min2021meta}, reconstruction and simulation modes are introduced in our training process. The processes of reconstruction~$\mathcal{L}_{recon}$ and simulation~$\mathcal{L}_{simu}$ can be described as:
\begin{equation}
\centering
\begin{split}
\mathcal{L}_{recon}=&\mathcal{L}_{mel}+(1-\alpha)*\mathcal{L}_{spk\_ce}+\alpha*\lambda_{spk}*\mathcal{L}_{spk\_cos}\\
&+\lambda_{c}*\mathcal{L}_{c}+\sum_{s}^{}\mathcal{L}_{style_{s}}+\lambda_{adv}\mathcal{L}_{adv}+\alpha*\mathcal{L}_{wReg},
\end{split}
\end{equation}
\begin{equation}
\centering
\begin{split}
\mathcal{L}_{simu}=&\lambda_{spk}*\mathcal{L}_{spk\_cos}+\lambda_{c}*\mathcal{L}_{c}+\sum_{s,s\ne l}^{}\mathcal{L}_{style_{s}}+\lambda_{adv}\mathcal{L}_{adv}
\end{split}
\end{equation}
where the value of $\alpha$ is 0 or 1 to indicate base model training and adaptation, respectively. In the simulation mode, we randomly select a source speech from different speakers to obtain content and style. Note that $\mathcal{L}_{mel}$ and $\mathcal{L}_{style_{l}}$ do not involve in this process without the ground-truth mel spectrum. With the high similarity between $h_{l}$ and mel spectrum, the $h_{l}$-based style loss $L_{style_{l}}$ is not considered in the simulation mode to avoid the effect on the speaker similarity. In practice, 
we first \textit{pre-train} the speaker classifier, speaker indicator, SER model, and content model. During the \textit{base model training}, the reconstruction mode is adopted to drive this process. When performing \textit{adaptation}, both reconstruction mode and simulation mode are utilized, which is helpful for the VC model’s generalization ability.

\vspace{-5pt}
\section{Experiments}

\label{sec:exp}
\subsection{Experimental Setup}
\subsubsection{Corpus}
Two hundred speakers of open-source Mandarin data Aishell3~\cite{aishell3} are used to train the base VC model. For low-resource testing, four reserved speakers of Aishell3 and four speakers of an internal dataset are selected as target speakers. 1 and 5 utterances of the target speaker are selected to conduct speaker adaptation, of which the total duration ranges from 3 to 4 seconds and 15 to 16 seconds. A test set contains a series of highly expressive speech in different scenarios, including emotions, movies, novels, and daily conversation. A set of 120 utterances is randomly selected for the test. The ASR model is an open-source model implemented by WeNet \renewcommand{\thefootnote}{\arabic{footnote}}toolkit
\footnote[1]{https://github.com/wenet-e2e/wenet} and trained with 10k hours of speech from Wenetpseech~\cite{zhang2022wenetspeech}. The SV model is an ECAPA-TDNN model~\cite{ECAPA_TDNN} trained with Voxceleb2~\cite{voxceleb2}. Open-source emotional dataset ESD~\cite{zhou2021seenESD} is used to train the SER model. Our universal TFGAN~\cite{tian2020tfgan} are trained on 1k hours of speech recordings containing 6k speakers. The content model and speaker indicator are both trained with Aishell3~\cite{aishell3}. 

\vspace{-3pt}
\subsubsection{Implement details}
All speech recordings are resampled to 24kHz and represented by 80-dim mel spectrum which is computed with 50ms frame length and 10ms frame shift. 
The conformer encoder consists of one conformer block following the official \renewcommand{\thefootnote}{\arabic{footnote}}implementation
\footnote[2]{https://github.com/sooftware/conformer}. In style module, reference encoder~\cite{ReferenceSkerryRyan2018Reference} and modified reference encoder~\cite{imLian2021TowardsFP} are based on the original model configuration. The mel decoder is an auto-regressive module~\cite{taco2shen2018natural}.
15 negative samples per step are used for speaker indicator pre-training. The down-sample rate in the style module and speaker indicator are both 20. In the base model training,
$\lambda_{c}$ and $\lambda_{adv}$ are set to $1$ and $0.05$, respectively. The discriminator is trained alternately every 1 step of training.
During adaptation, 
$\lambda_{spk}$, $\lambda_{c}$, and $\lambda_{adv}$ are set to $0.1$, $0.1$, and $0.05$, respectively. The VC model is trained for $400$ epochs, in which process the learning rate starts from $5*10^{-5}$ and decays every $30$ epochs with a decay rate of $0.5$. Note that, during adaptation, reconstruction and simulation modes are simultaneously activated. When adapting to a target speaker with only 1 or 5 utterances, 10 and 25 sentences are randomly selected for simulation at each step.

\vspace{-10pt}
\subsection{Experimental Results}
To evaluate the performance of the proposed MFC-StyleVC on the low-resource VC task, two recently proposed VC systems, i.e., \textbf{Hybrid-VC}~\cite{imwang2021enriching} and \textbf{SN-VC}~\cite{wang2022one}, are compared in the experiments. Hybrid-VC is the state-of-the-art method on delivering speaking style without further adaptation, while SN-VC is the state-of-the-art method with speaker adaptation. Besides, a variant of MFC-StyleVC, referred to as \textbf{StyleVC}, without multi-factor constraint module, is also compared. The performance of different models is evaluated both subjectively and objectively. Fifteen participants in total joined in the listening tests. We highly recommend readers listen to the converted samples from the demo
\renewcommand{\thefootnote}{\arabic{footnote}}page
\footnote[3]{Samples can be found in {\url{https://kerwinchao.github.io/lowresourcevc.github.io/}}\label{demo}}.

The results are shown in Table~\ref{exp:mos&obj}, in which \textit{Speech Quality} and \textit{Speaker Similarity} are subjective evaluation metrics that indicate the mean opinion scores (MOS) in terms of speech quality and speaker similarity, respectively. The value of MOS is shown with confidence intervals of $95\%$. Five utterances are randomly selected from the test set for each target speaker (5 * 8 = 40) in the subjective test. The character error rate (CER) is obtained by an ASR model and is to present the speech intelligible of converted speech. Lower CER means better speech intelligibility. For style similarity, a comparative mean opinion score (CMOS) test is performed between the comparison and proposed systems. A positive CMOS value means that the compared method has better style consistency with source speech than the proposed method and vice versa. Meanwhile, $\mathcal{D}_{style}$ measures the distance between the style embeddings of source speech and converted speech, which is the sum of the $\mathcal{L}_{style_{m}}$ and $\mathcal{L}_{style_{h}}$. With containing rich speaker-related information, $\mathcal{L}_{style_{l}}$ is not considered. $P_{lf0}$ is the Pearson correlation coefficients of lf0 between the source speech and converted speech. Both $\mathcal{D}_{style}$ and $P_{lf0}$ are used to show the performance of VC models on the style delivery. Lower $\mathcal{D}_{style}$ and higher $P_{lf0}$ means better performance. Cosine similarity (Cos.Sim) between the SV model-based speaker embeddings of converted speech and speech from the target speaker shows speaker timbre similarity between them.

As shown in Table~\ref{exp:mos&obj}, the number of utterances from the target speaker shows an obvious effect on the VC performances of all models. In general, the longer the duration, the better the results, which is further illuminated in our page\textsuperscript{\ref{demo}}. To be specific, all models' performance degrades when the utterance number reduces to 1 from 5. The proposed method achieves the best results for speaker similarity and speech quality on both 1 and 5 utterances. Meanwhile, MFC-StyleVC achieves the best CER scores and higher Cos.Sim. It shows that the proposed method can effectively learn low-resource speaker timbre and maintain the high quality of the converted speech. From the CMOS results, it is found that all the values are less than 0, which indicates that the proposed model is effective in conveying style information under low-resource tasks. The $D_{style}$ and $P_{lf0}$ of different models also present similar results. Compared with StyleVC, MFC-StyleVC with multi-factor constraints is significantly improved in all three aspects. Moreover, Hybrid-VC maintains a high style similarity, while training on limited data exacerbates the speaker timbre of source speech leaking to converted speech and leads to poor speaker similarity. Contrary to HybridVC, SN-VC can achieve relatively good speaker similarity but suffers from poor style similarity due to poor ability of style modeling and speaker normalization on a highly expressive testset.

\vspace{-5pt}
\subsection{Effectiveness Verification}
\subsubsection{Ablation Analysis}
Table~\ref{exp:ablation} shows the effect of different components of MFC-StyleVC on the VC task. As can be seen, the constraints of style, content, and speaker play important roles in MFC-StyleVC. The dropping of these constraints brings obvious performance decreases in terms of style similarity ($D_{style}$), speech intelligible (CER), and speaker similarity, respectively. In addition to constraints mentioned above, the adversarial optimization method based on the generated spectrum is also analyzed in this table. As shown, when the model is trained without the adversarial loss, the performance drops in terms of all evaluation metrics, indicating the effectiveness of this optimization method. Furthermore, to alleviate the mismatch between training and inference, the simulation mode is introduced in the training process. When this simulation mode is excluded from the training process, obvious performance decrease appears in the CER and $D_{style}$, demonstrating the importance of this simulation training method.

\begin{table}[]
\scriptsize
\centering
\caption{Ablation analysis of different components and training methods.}
\vspace{3pt}
\label{exp:ablation}
\begin{tabular}{c|cc|cc|cc}
\hline
               & \multicolumn{2}{c|}{CER $(\downarrow)$}         & \multicolumn{2}{c|}{$D_{style} (\downarrow)$}   & \multicolumn{2}{c}{Cos.Sim. $(\uparrow)$}       \\ \hline
Length          & \multicolumn{1}{c|}{1 utt.} & 5 utt. & \multicolumn{1}{c|}{1 utt.} & 5 utt. & \multicolumn{1}{c|}{1 utt.}  & 5 utt.  \\ \hline
MFC-StyleVC  & \multicolumn{1}{c|}{\textbf{3.98}} & \textbf{3.74} & \multicolumn{1}{c|}{\textbf{2.26}} & 2.16 & \multicolumn{1}{c|}{0.804} & \textbf{0.828} \\ 
w/o $L_{style}$   & \multicolumn{1}{c|}{4.20} & 3.98 & \multicolumn{1}{c|}{2.65} & 2.58 & \multicolumn{1}{c|}{\textbf{0.812}} & 0.815 \\
w/o $L_{content}$ & \multicolumn{1}{c|}{4.49} & 4.20 & \multicolumn{1}{c|}{2.51} & 2.23 & \multicolumn{1}{c|}{0.809} & 0.826 \\
w/o $L_{spk}$     & \multicolumn{1}{c|}{4.20} & 4.05 & \multicolumn{1}{c|}{2.46} & \textbf{2.14} & \multicolumn{1}{c|}{0.800} & 0.810 \\
w/o $L_{adv}$     & \multicolumn{1}{c|}{\textbf{3.98}} & 3.84 & \multicolumn{1}{c|}{2.59} & 2.34 & \multicolumn{1}{c|}{0.803} & 0.813 \\
w/o Simulation  & \multicolumn{1}{c|}{4.56} & 3.91 & \multicolumn{1}{c|}{3.09} & 2.73 & \multicolumn{1}{c|}{0.808} & 0.824 \\ 
\hline
\end{tabular}
\vspace{-10pt}
\end{table}

\begin{table}[ht]
\vspace{-10pt}
\centering
\scriptsize
\setlength{\tabcolsep}{2mm}
\caption{Intra and inter speaker cosine similarity }
\vspace{3pt}
\label{exp:speaker_model}
\begin{tabular}{c|c|c|c}
\hline
      & Speaker Classifier & Speaker Verification & Speaker Indicator \\ \hline
Intra $(\uparrow)$ & 0.90               & 0.89                 & \textbf{0.96}              \\ \hline
Inter $(\downarrow)$ & 0.72               & 0.33                 & \textbf{0.13}              \\ \hline

\end{tabular}
\vspace{-5pt}
\end{table}

\vspace{-10pt}
\subsubsection{Comparison of different speaker models}
In MFC-StyleVC, the speaker indicator, which is optimized with a contrastive learning method, is proposed for the speaker timbre constraint. Here, we would like to show the performance of the speaker indicator in clustering speakers. The quantified performances are shown in Table~\ref{exp:speaker_model}, in which the cosine similarity of inter and intra speakers is presented. As can be seen, speaker embeddings obtained by the SV model and speaker indicator present a similar performance in terms of inter-speaker similarity, which is obviously better than that achieved by the speaker classifier. However, the intra-speaker similarity achieved by the SV model does not outperform that obtained by the classifier model and even gets worse performance. In contrast, the speaker indicator model gets higher intra-speaker similarity and also lower inter-speaker similarity, indicating the good design of the speaker indicator.

\section{CONCLUSION}

In this work, a novel VC framework MFC-StyleVC designed for low-resource VC task that considers the source speaker's speaking style is proposed. To alleviate the over-fitting issue in the low-resource case, multi-factor constraints are proposed in MFC-StyleVC to explicitly constrain different aspects of converted speech. Besides, a simulation training mode, which simulates the inference process during the training, is introduced. Experiments conducted on the VC task have demonstrated that the proposed method achieves new state-of-the-art performance on the low-resource VC task.



\bibliographystyle{IEEE}
\footnotesize
\bibliography{strings,refs}

\end{document}